\documentclass[conference]{IEEEtran}
\IEEEoverridecommandlockouts
\usepackage{cite}
\usepackage{makecell}

\usepackage{amsmath,amssymb,amsfonts}
\usepackage{algorithmic}
\usepackage{graphicx}
\usepackage{textcomp}
\usepackage{xcolor}
\def\BibTeX{{\rm B\kern-.05em{\sc i\kern-.025em b}\kern-.08em
    T\kern-.1667em\lower.7ex\hbox{E}\kern-.125emX}}
\begin{document}

\title{FPDANet: A Multi-Section Classification Model for Intelligent Screening of Fetal Ultrasound}

\author{
\IEEEauthorblockN{
Minglang Chen\IEEEauthorrefmark{1},
Jie He\IEEEauthorrefmark{2}\thanks{*Corresponding author: Jie He (e-mail: hejie1213@hnu.edu.cn).},
Caixu Xu\IEEEauthorrefmark{1}, 
Bocheng Liang\IEEEauthorrefmark{3}, 
Shengli Li\IEEEauthorrefmark{3}, 
Guannan He\IEEEauthorrefmark{4}, 
and Xiongjie Tao\IEEEauthorrefmark{5}}
\IEEEauthorblockA{\IEEEauthorrefmark{1}(a) Guangxi Key Laboratory of Machine Vision and Intelligent Control, Wuzhou University, Wuzhou, 543002 China}
\IEEEauthorblockA{\IEEEauthorrefmark{2}(b) College of Computer Science and Electronic Engineering, Hunan University, Changsha, 410082 China}
\IEEEauthorblockA{\IEEEauthorrefmark{3}(c) Department of Ultrasound, Shenzhen Maternal and Child Healthcare Hospital, Shenzhen, 518100 China}
\IEEEauthorblockA{\IEEEauthorrefmark{4}(d) Department of Ultrasound, Sichuan Provincial Maternity and Child Healthcare Hospital, Chengdu, 610100 China}
\IEEEauthorblockA{\IEEEauthorrefmark{5}(e) Faculty of Humanities and Arts, Macau University of Science and Technology, Macau, 999078 China}
\thanks{This work was supported in part by the National Natural Science Foundation of China under Grants 62162054, Basic Ability Improvement Project for Young and Middle-aged Teachers in Guangxi, China under Grants 2024KY0694 and 2023C004, Science and Technology Development Fund, Macau SAR (0009/2024/ITP1), Open Fundation of Guangxi Key Laboratory of Machine Vision and Intelligent Control under Grants 2022B06 and Wuzhou science and technology plan project under Grants 202302036.}
}

\maketitle

\begin{abstract}
ResNet has been widely used in image classification tasks due to its ability to model the residual dependence of constant mappings for linear computation. However, the ResNet method adopts a unidirectional transfer of features and lacks an effective method to correlate contextual information, which is not effective in classifying fetal ultrasound images in the classification task, and fetal ultrasound images have problems such as low contrast, high similarity, and high noise. Therefore, we propose a bilateral multi-scale information fusion network-based FPDANet to address the above challenges. Specifically, we design the positional attention mechanism (DAN) module, which utilizes the similarity of features to establish the dependency of different spatial positional features and enhance the feature representation. In addition, we design a bilateral multi-scale (FPAN) information fusion module to capture contextual and global feature dependencies at different feature scales, thereby further improving the model representation. FPDANet classification results obtained 91.05\% and 100\% in Top-1 and Top-5 metrics, respectively, and the experimental results proved the effectiveness and robustness of FPDANet.
\end{abstract}

\begin{IEEEkeywords}
Ultrasound screening during pregnancy, image classification, machine vision, deep learning
\end{IEEEkeywords}

\begin{figure*}[tbp]
	\centering   			
    \includegraphics[width=1\textwidth]{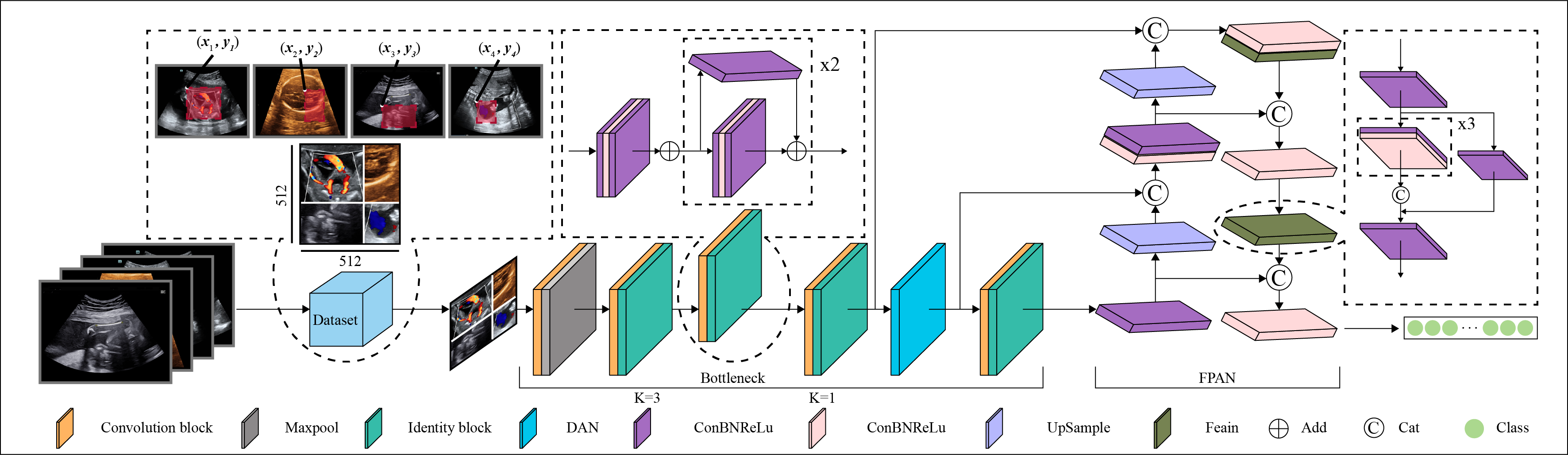}
	\caption{FPDANet ultrasound image classification model.}
	\label{fig1}
 \vspace{-5mm}
\end{figure*}

\section{Introduction}
Prenatal fetal screening is an effective means of screening for birth defects, which includes ultrasonography and magnetic resonance imaging~\cite{gangning}, and ultrasonography has become the primary modality for prenatal fetal health screening because of its non-destructive, inexpensive, and safe properties~\cite{sun2021prenatal}. However, fetal ultrasound images suffer from similar anatomical structures and morphology, low contrast, and interference from noise and artifacts~\cite{gudigar2022role}. Moreover, according to our prenatal ultrasound guidelines for screening during pregnancy~\cite{lishenglli}, for the task of fetal cardiac ultrasound screening alone, the physician needs to examine 5 regions and obtain at least 22 standard views and perform more than 20 evaluations, which involves approximately 52 views throughout the entire gestation~\cite{crino2013aium,salomon2011practice}. This makes it time-consuming and laborious for doctors to select the sections~\cite{gangning}, and the correctness and authority of section selection depends on the doctor's clinical experience and professionalism, which in turn increases the difficulty of fetal prenatal screening. Deep learning methods are utilized to automatically and rapidly select and differentiate ultrasound images from different parts of the body in order to assist ultrasonographers in the next step of screening and diagnosis, to alleviate the work pressure of ultrasonographers, and to reduce the subjectivity of section selection, thus contributing to the enhancement of medical intelligence.

Deep learning has been widely used in the field of medical pathology diagnosis and also provides a new solution idea for the task of intelligent prenatal screening of fetuses. However, existing deep learning methods in the medical field have the following challenges. (1) Existing deep learning methods are based on natural image design with a single method of feature extraction and delivery ~\cite{sarwinda2021deep,li2018automatic,pu2021fetal}, which is difficult to adapt to the task of fetal ultrasound image classification with low contrast and complexity. (2) Prenatal fetal examination is not limited to four-chamber cardiac views but also includes 21 types of views, such as ventricles, and includes imaging modalities such as color Doppler flow imaging (CDFI) spectral Doppler, and two-dimensional ultrasound images~\cite{gangning}, which provide complex and varied image information and cause serious interference with the extraction of model features~\cite{gao2023deep,pu2022mobileunet,lu2022yolox}. In order to cope with the above challenges, this paper designs a Fetal Prenatal Intelligent Screening Model (FPDANet) based on a dual-head feature enhancement strategy. FPDANet contains backbone networks, bilateral attention mechanisms and dual-head feature enhancement modules.

\section{RELATED WORK}
Clinically experienced physicians are highly specific and sensitive to ultrasound imaging, but there is still much room for improvement in the diagnostic efficiency of fetal prenatal screening in remote areas and other regions that lack highly qualified medical personnel. Deep learning methods will not only capture critical and unique feature information, but they will also allow for remote-assisted diagnosis. Currently, the mainstream image classification methods such as VGGNet, ResNet, etc., feature transfer are unidirectional. The features have strong singularity, and it is easy to have problems such as difficult model fitting when there are more classification tasks or deeper model depth. He~\cite{he2024fetal} et al. proposed a multi-task learning model for identification and localization of key anatomical structures and quality control of standard cuts in fetal cardiac ultrasound images based on YOLOv5, which enhances the robustness of the cuts identification through a multi-scale feature fusion module. Pu~\cite{pu2021automatic} et al. proposed a multi-task standard ultrasound section recognition model based on the spatio-temporal characteristics of ultrasound video streams. The convolutional neural network of the model is used to recognize the key spatial feature information of fetal ultrasound images in each frame of the video and learn the spatial features in the standard ultrasound section, and the two fusion strategies are introduced to improve the robustness of the model in the task of recognizing the standard section. Zhao~\cite{zhao2022ultrasound} et al. come out with an ultrasound-standardized planar detection model based on multi-task learning and hybrid knowledge graphs. The model learns shared features of fetal ultrasound images through convolutional blocks and extends the shared features to task-specific output streams to optimize the generalization performance of the model.

\section{METHOD}

\subsection{FPDANet Overview}
Fetal ultrasound images also have important differences from traditional transmission images. Fetal ultrasound images mainly focus on fetal anatomy, physiological functions, and pathological changes, and ultrasound imaging is susceptible to the interference of a variety of internal and external factors~\cite{pu2024hfsccd}, such as different fetal development cycles~\cite{puunsupervised,chen2020sleep,pang2024adaptive}, scanning area equipment~\cite{pu2024m3} and equipment parameters~\cite{zhao2024farn,pang2022efficient}, and imaging in terms of resolution, contrast, and noise level, which leads to the model in the feature extraction and understanding of the difficulty of capturing generic feature representations, which in turn leads to poor classification results, especially in multi-category ultrasound image classification tasks, where generic features are more difficult to capture.

Image classification needs to focus on the global content information of the image. For multi-category ultrasound image classification, the model suffers from complexity and sparsity in the extracted features. Therefore, deeper networks and multiple neurons are required for better understanding, yet deeper networks are prone to loss explosion and gradient vanishing~\cite{sarwinda2021deep}. For this reason, this paper uses the ResNet model as the backbone network, which can deepen the network and avoid the problem of loss explosion and gradient disappearance, theoretically superimposed multi-layer network layers to make the features have a larger sense of the wild~\cite{zhao2023transfsm,he2024fetalx,li2024shift}. However, ResNet is a unidirectional feature-passing model, which tends to ignore contextual information and thus has difficulty converging in classification tasks.

In summary, this paper proposes a convolutional neural network for automatic classification of multi-type fetal ultrasound images by using ResNet model as the backbone network and the bilateral multi-scale information fusion network levy enhancement module, as shown in Fig.~\ref{fig1}. The model is mainly divided into three core components, the backbone network, attention mechanism, and bilateral multi-scale information fusion network. The underlying detail information of the image is extracted through the backbone network, while the feature multi-scale fusion method of the bilateral multi-scale information fusion network is utilized to fuse the underlying detail information with the high-level semantic information to enhance the expression of rich semantic and detail information of the features so as to improve the accuracy and generalization ability of the model classification.

In deep learning networks, the deeper the features have, the richer the semantic information become. However, the deeper the model architecture, the more prone it is to problems such as gradient disappearance and explosion. Therefore, in order to obtain semantically rich feature information and prevent gradient disappearance at the same time, in this paper, we take the convolution block and the identity block as the basic components of the backbone network.

In the backbone network, the underlying feature extraction layer consists of a convolution block and an identity block in series. Convolution block The main branch contains $3\times 3$ convolutional layers and two $1\times 1$ convolutional layers, which are used for feature extraction, dimensionality reduction, or dimensionality upgrading, respectively, and the side branch is fused with the main branch after dimensionality reduction with $1\times 1$ convolution, and the output features of each layer are batch normalized to increase the nonlinear expression. Identity blocks carry out constant feature transfer, which is used for feature transfer and abstract feature mining. The difference with the convolution block is that the last layer of $1\times 1$ convolution in the main branch is upscaled to the input feature scale, and the side branch outputs the original features. It improves model performance and stability, broadens network depth, and mitigates the gradient vanishing problem without adding too many parameters and computational effort.

\subsection{DAN Attention Network}
The backbone network consists of Convolution block and Identity block superimposed, Identity block is a feature short-circuit connection, the side branch and the main branch to enhance the feature direct integration, alleviate the problem of gradient disappearance, but due to the short-circuit to the outliers, noise propagation, so that the model is sensitive to the impact of the classification performance, in this paper, the bottom of the layer to the high level of feature enhancement, pre-use attention mechanism to enhance the core information, inhibit the interference features, and improve the classification performance of the model.

\begin{figure}[htbp]
	\centering   			
    \includegraphics[width=0.5\textwidth]{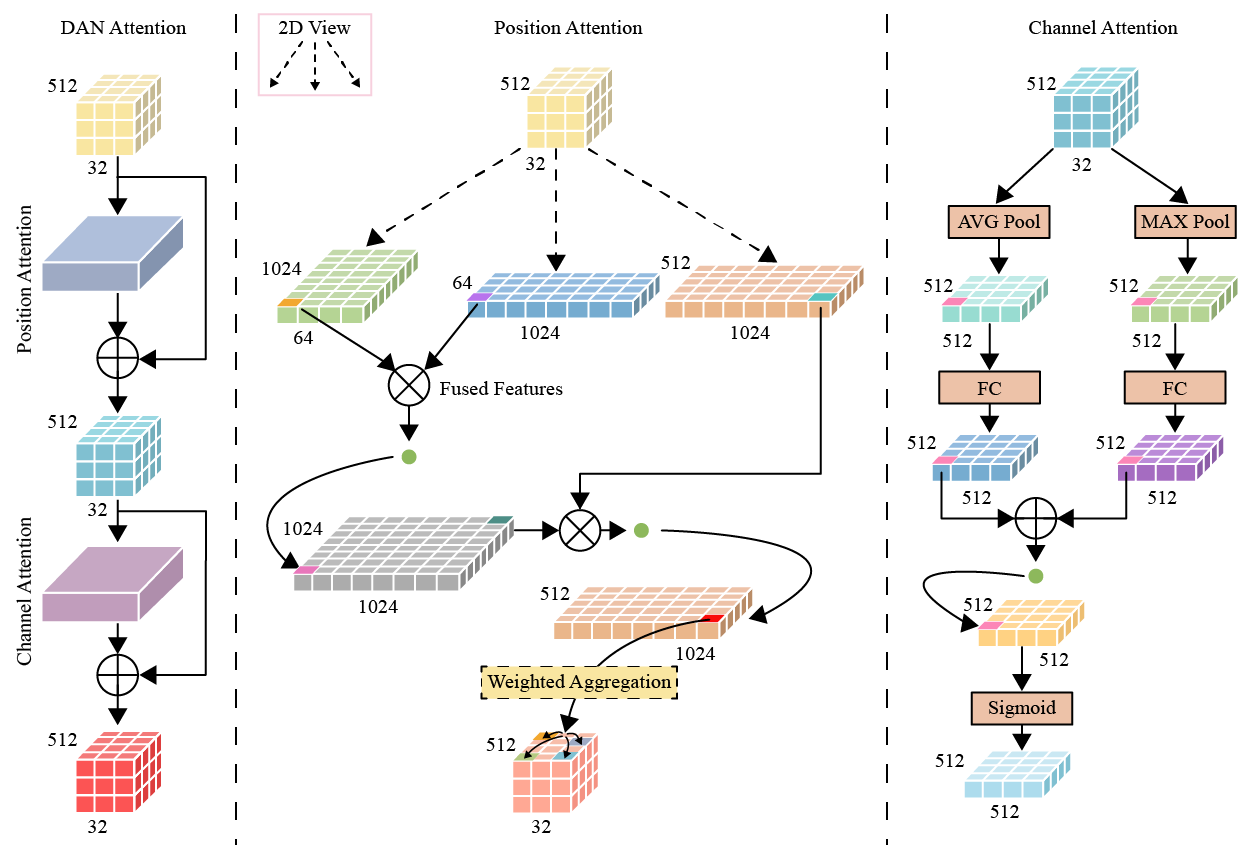}
    \vspace{-2mm}
	\caption{Bilateral attention mechanisms.}
	\label{DAN}
    
\end{figure}

The backbone network consists of five groups of residual groups, the fourth group of residual block output features with rich high-level semantic information, with a high degree of abstraction, storage of local details, the attention mechanism to help capture the feature relationships, the accurate allocation of weights, this paper in the backbone network of the fourth and the fifth group of residual groups, the introduction of parallel location and channel attention module composed of the DANet, in order to enrich the feature information, as shown in Fig.~\ref{DAN}.
\vspace{-4mm}

\begin{table*}
    \centering
    \setlength{\tabcolsep}{0.9pt}
    \renewcommand{\arraystretch}{1.2}
    \caption{Image type and data distribution for training, test and validation sets.}
    
    \resizebox{\linewidth}{!}{
    \begin{tabular}{c|ccccccccccccccccccccc|c}
    \hline
        Item & 3VT & BL & 4C & TV & UR & HL & ICO & FL & CTSP & TF & CI & TTV & DI & Ab & LVOT & Kidneys & Eyes & TCV & MFP & LSSP & RVOT & All\\ 
        Train & 309 & 330 & 277 & 329 & 328 & 329 & 330 & 329 & 330 & 328 & 331 & 330 & 185 & 329 & 288 & 308 & 308 & 315 & 316 & 312 & 313 & 6554 \\ 
        Val & 77 & 82 & 69 & 82 & 82 & 82 & 82 & 82 & 82 & 82 & 82 & 82 & 46 & 82 & 71 & 77 & 76 & 78 & 78 & 77 & 78 & 1629 \\ 
        Test & 43 & 46 & 39 & 46 & 46 & 46 & 46 & 46 & 46 & 46 & 46 & 46 & 26 & 46 & 40 & 43 & 43 & 44 & 44 & 44 & 44 & 916 \\ \hline
    \end{tabular}
    }
    \label{table1}
    \vspace{-5mm}
\end{table*}
\vspace{-3mm}
The positional attention module establishes the dependence of similar features at different spatial locations of an image, allowing the model to understand the global structure and contextual information. The input feature \(A\in\mathbb{R^{C\times H \times W}}\)  is first flattened into vector B, C, D, \(\{B, C, D\} \in\mathbb{R^{C\times H \times W}}\) by 1×1 convolution, then features B and C are transformed into \(\mathbb{R^{C\times N}}\) shape, where \(N = H \cdot W \) is the number of pixels, and then features B and C are multiplied by the transposition matrix, and (\(S\in\mathbb{R^{C\times N}}\)) is calculated by softmax.
\begin{align} 
S_{ji}=\frac{exp(B_i \cdot C_j)}{\sum_{i=0}^{N} exp(B_i \cdot C_j)},
\tag{1} 
\label{eq1}
\end{align}
\(S_{ji}\) describes the effect of position \(i^{th}\) on position \(j^{th}\). The more similar the features, the greater the correlation. Finally, multiply the transpose matrix of feature map D and feature map S and multiply by the scale parameter $\alpha$.
\begin{align} 
E_{j}=\alpha \sum_{i=1}^{N} (S_{ji}D_i)+A_j,
\tag{2} 
\label{eq2}
\end{align}
where $\alpha$ is initialized to 0 and gradually learns to assign more weights. From equation \ref{eq2}, it can be seen that the result of each position in E is a weighted sum of the features and original features of all positions; thus, it has a global view of the context and selectively aggregates contexts according to the spatial attention map. Similar semantic features are reinforced with each other, thus enhancing intra-class compactness and semantic consistency.

The channel attention mechanism directly reshapes the input feature A into \(\mathbb{R^{C \cdot C}}\), then multiplies the feature A with its transposed feature matrix, and finally computes the channel attention map \(X\in\mathbb{R^{C \cdot C}}\) by passing it through softmax.
\begin{align} 
X_{ji}=\frac{exp(A_i \cdot A_j)}{\sum_{i=1}^{C} exp(A_i \cdot A_j)},
\tag{3} 
\label{eq3}
\end{align}
where \(X_{ji}\) describes the effect of the \(i^{th}\) position on the \(j^{th}\) position. In addition, a matrix multiplication is performed on the transpose of X and A, and the result reshapes into \(\mathbb{R^{C \cdot H \cdot W}}\). Finally, a scaling factor $\beta$ is multiplied and summed element-wise with A to obtain the final output feature \(E\in\mathbb{R^{C \cdot H \cdot W}}\).
\begin{align} 
E_{j}=\beta \sum_{i=1}^{C} (X_{ji}A_i)+A_j,
\tag{4} 
\label{eq4}
\end{align}
where $\beta$ learns a weight gradually from 0. Equation \ref{eq4} shows that the final feature of each channel is a weighted sum of the features of all channels and the original feature, which models the long-range semantic dependencies between feature maps and helps to improve feature discrimination.

\subsection{FPAN Bilateral Multi-scale Information Fusion Network}
There exists local feature multi-scale fusion within the backbone network, but the overall architecture is feature unidirectional conduction, and its extracted features lack of upper and lower related information and low richness. In this paper, a feature enhancement network for bilateral multi-scale information fusion is designed using bottom-up and top-down tandem design architectures. The bottom-up network can transmit high-level features to the bottom layer to help it understand large-scale target features. The top-down network can pass the bottom-up features to a higher level to solve the problem of small target with few pixels and the loss of feature information, and the lower layer of high-resolution feature maps contains important detail information, allowing it to better describe the small target features.

FPAN network takes the last layer of features of the backbone network as the input of the bottom-up network, fuses the features of the fourth and third layers of the backbone network through the upper sampling layer, takes the topmost layer of features as the input of the top-down network, fuses the features of the bottom-sampled feature layer with the output features of the bottom layer of the bottom-up network and the intermediate layer of the bottom-up network, and introduces convolutional layers after the upper and lower sampling layers to eliminate the effect of feature aliasing. Through the FPAN network, different levels of features are fused, which helps the model to extract and understand the target feature information at different scales.

\section{EXPERIMENTS}
\subsection{Dataset}
We collected a dataset of 21 fetal ultrasound images containing 2D ultrasound images, CDFI and spectral Doppler from Shenzhen Maternal and Child Health Hospital, which was acquired between 14 and 28 weeks of gestation using equipment such as Philips and Siemens, and a total of 9099 images were collected. The dataset is divided into a training dataset, a validation dataset, and a test dataset in the ratio of 7:2:1 as shown in Table~\ref{table1}. The dataset is divided into training dataset, validation dataset, and test dataset. The names as well as abbreviations are shown in Table~\ref{tablen2}. Informed consent was obtained from all patients, and the study was approved by the Ethics Committee of Shenzhen Maternal and Child Health Hospital.
\begin{table*}
    \centering
    \setlength{\tabcolsep}{0.9pt}
    \renewcommand{\arraystretch}{1.2}
    \caption{Abbreviation of the medical specialty corresponding to the full name of the section.}
    \resizebox{\linewidth}{!}{
    \begin{tabular}{c|ccccccc}
    \hline
        Full name & {\makecell{Three vessel\\tracheal}} & {\makecell{Bladder\\Axial View}} & {\makecell{Four-Chamber\\Viewr}} & {\makecell{Transventricular\\View}} & {\makecell{Ulna and Radius\\Coronal View}} & {\makecell{Humerus Long\\Axis View}} & {\makecell{Internal Cervical Os\\Sagittal View}} \\ 
        
        Abbreviation & 3VT & BL & 4C & TV & UR & HL & ICO \\ \hline

        Full name  & {\makecell{Femur Long\\Axis View}} & {\makecell{Cervicothoracic Spine\\Sagittal View}} & {\makecell{Tibia and Fibula\\Coronal View}} & {\makecell{Cord Insertion\\Abdominal Axial View}} & {\makecell{Tranthalamic\\View}} & {\makecell{Diaphragm\\Coronal View}} & {\makecell{Upper Abdomen\\Axial View}} \\
        Abbreviation & FL & CTSP & TF & CI & TTV & DI & Ab \\ \hline
        
        Full name & {\makecell{Left ventricular\\outflow tract}}& {\makecell{Kidneys Axial\\View}} & {\makecell{Eye Axial\\View}} & {\makecell{Transcerebellar\\View}} & {\makecell{Median Sagittal Facial\\Profile View}} & {\makecell{Lumbosacral Spine\\Sagittal View}} & {\makecell{Right ventricular\\outflow tract}}\\ 
        Abbreviation & LVOT & Kidneys & Eyes & TCV & MFP & LSSP & RVOT\\ \hline
    \end{tabular}
    }
    \label{tablen2}
    \vspace{-5mm}
\end{table*}
\vspace{-5mm}

\begin{table*}
\centering
\setlength{\tabcolsep}{0.9pt}
\renewcommand{\arraystretch}{1.5}

\caption{Effects of different attention mechanisms on feature extraction.}
\resizebox{\linewidth}{!}{
\begin{tabular}{c|ccccccccccccccccccccc|cc}
  \hline
Item & 3VT & BL & 4C & TV & UR & HL & ICO & FL & CTSP & TF & CI & TTV & DI & Ab & LVOT & Kidneys & Eyes & TCV & MFP & LSSP & RVOT & TOP1 (\%) & TOP5 (\%)\\
 \hline
None & \textbf{1.00} &  \underline{0.90} & 0.88 & 0.70 &  \underline{0.77} & 0.79  & 0.93 & 0.86 &  \underline{0.98} & 0.63 & 0.86 & \underline{0.71} & \underline{0.96} & \underline{0.94} & 0.79 & 0.91 & 0.96 & \underline{0.98} & \underline{0.96} & \underline{0.96} & \textbf{0.98} & 87.12 & 99.67 \\
CBAM & \textbf{1.00} & \textbf{0.92}  & 0.90 &  \underline{0.77} & \textbf{0.81} & 0.79 & 0.92 & 0.81 &  \underline{0.98} & 0.75 & \textbf{0.90} & 0.67 & \underline{0.96} & 0.88 & \textbf{0.83} & \textbf{1.00} & \textbf{1.00} & \underline{0.98} & \textbf{0.98} & \textbf{0.98} & \underline{0.90} & 88.76 & 99.67 \\

GAM  & \underline{0.98} & 0.88 & 0.84 & 0.59 & 0.66 &  \underline{0.80} & \textbf{0.96} & 0.80 &  \underline{0.98} & 0.66 & 0.84 & 0.62 & 0.91 & 0.92 & 0.79 & 0.91 & 0.95 & \underline{0.98} & 0.93 & \textbf{0.98} & 0.88 & 84.61 & 99.78 \\
    
CA  & \textbf{1.00} & \textbf{0.92} &  \underline{0.92} & 0.76 & 0.73 & \textbf{0.82} & 0.94 & \textbf{0.93} & 0.96 &  \underline{0.78} & 0.80 & 0.69 & 0.88 & 0.85 & \underline{0.81} & \underline{0.97} & \textbf{1.00} & 0.96 & \underline{0.96} & \textbf{0.98} & \underline{0.90} & \underline{87.99} & \underline{99.98} \\ 
                           
Ours & \textbf{1.00} &  \underline{0.90} & \textbf{0.95} & \textbf{0.80} & \textbf{0.81} & \textbf{0.82} &  \underline{0.95} &  \underline{0.91} & \textbf{1.00} & \textbf{0.79} & \underline{0.88} & \textbf{0.85} & \textbf{1.00} & \textbf{0.98} & \underline{0.81} & 0.93 & \underline{0.98} & \textbf{1.00} & \textbf{0.98} & \textbf{0.98} & 0.86  & \textbf{91.05} & \textbf{1.00}  \\
                       
\hline
\end{tabular}
}
\label{table3}

 \vspace{-5mm}
\end{table*}

\vspace{+3mm}
\subsection{Implementation Details}
 All experiments were done on an Arch Linux system with an Intel Xeon Platinum 8360 Y CPU and 4×NVIDIA A100 GPU using PyTorch. The optimizer for loss in this article is Adamw~\cite{zhuang2022understanding}, the learning rate decreases in steps, and the loss calculation method is CrossEntropyLoss~\cite{zhang2018generalized}. The maximum and minimum learning rates for initialization are 0.01 and 0.0001, respectively. The upper and lower limits of the learning rate are 0.001 and 0.0001, respectively. In order to dynamically adjust the learning rate and improve the stability of the model, we can adaptively adjust the maximum and minimum learning rates based on batch size. The calculation formula is as follows,
 \begin{align} 
lr_{mid}\_max=max(\frac{batch\_size}{nbs \cdot lr_{max}\_init},lr_{min}\_lim),
\tag{5} 
\label{eq5}
\end{align}
\vspace{-8mm}

\begin{align} 
lr_{max}=min(lr_{mid}\_max,lr_{max}\_lim).
\tag{6} 
\label{eq6}
\end{align}
\vspace{-8mm}

\begin{align} 
lr_{mid}\_min=max(\frac{batch\_size}{nbs \cdot lr_{min}\_init},\frac{lr_{min}\_lim} {100}),
\tag{7} 
\label{eq7}
\end{align}
\vspace{-8mm}

\begin{align} 
lr_{min}=min(lr_{mid}\_min,\frac{lr_{max}\_lim} {100}),
\tag{8} 
\label{eq8}
\end{align}
 the value of nbs is 64, and the model has been trained for a total of 200 epoch.

\subsection{Attention Mechanism Ablation Experiment}
In order to fully verify the effects of different attention mechanisms on the performance of the multi-category fetal ultrasound image classification model proposed in this paper, extensive experiments on the ablation of attention mechanisms were carried out under the same configuration parameters of the backbone network and FPAN network, as shown in Table~\ref{table3}. In the backbone network, mainstream attention mechanism modules such as CBAM, CA, and GAM\cite{liu2021global} are introduced, respectively, and None indicates that no attention mechanism is introduced. The experimental results show that the introduction of a reasonable attention mechanism can effectively improve the classification performance of the model, especially the DAN attention mechanism introduced in this paper, which improves the evaluation index by 3.93\% compared with the baseline model Top-1\cite{xu1997conference}. And the result of GAM attention mechanism is lower than the original model because GAM attention mechanism is a global attention mechanism, which has some ambiguity in explaining the decision-making process of the model. Especially in the image classification task where the image content is complex and there are multiple interrelated objects, the GAM attention mechanism assigns attention weights to each region of the whole image, and the information of the whole image is considered in the weight calculation, which leads to the calculation of the weights being affected by the interaction of many factors, and then it is difficult to make the weight decision for the main content exactly.


\begin{figure*}[htbp]
	\centering   			
    \includegraphics[width=1\textwidth]{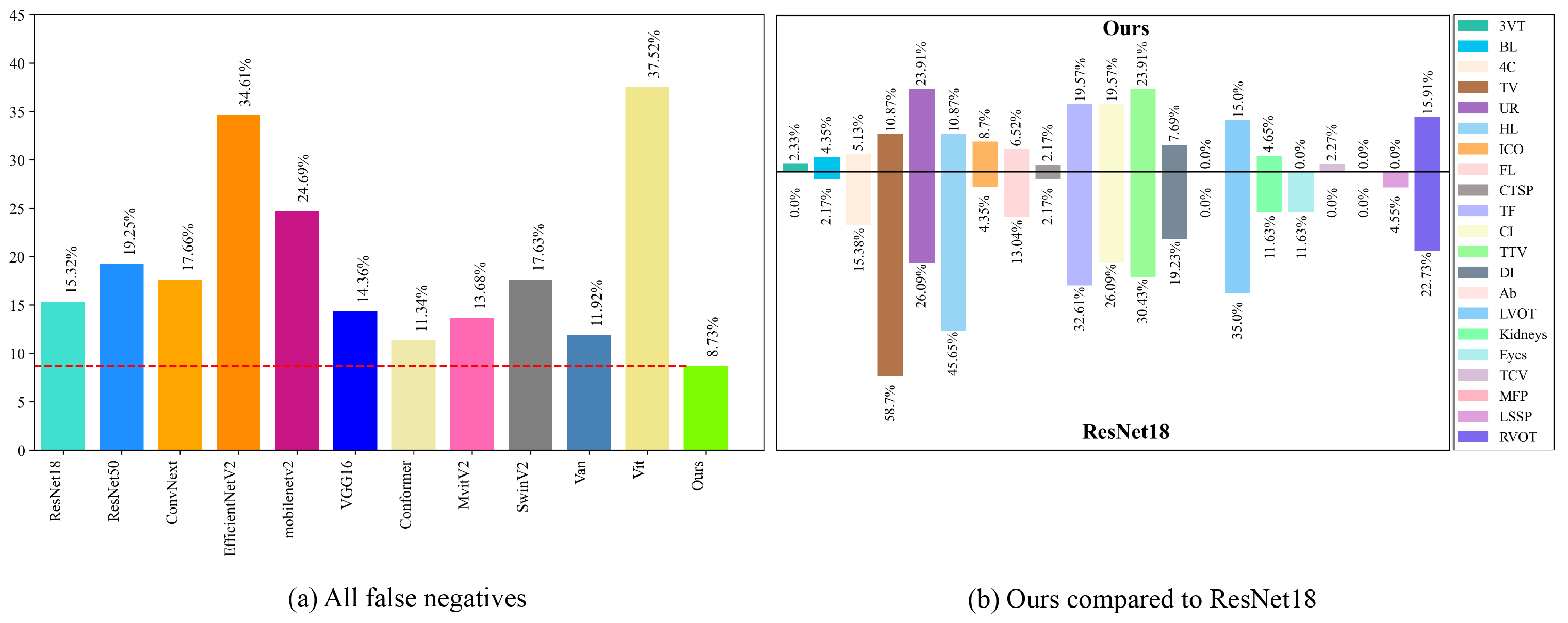}
	\caption{False negative rate curves under different baseline models.}
	\label{FNRs}
\end{figure*}


\begin{table*}
    \centering
    \setlength{\tabcolsep}{1pt}
    \renewcommand{\arraystretch}{1.5}
    \caption{Classification effects of different baseline models.}
    \resizebox{\linewidth}{!}{
    \begin{tabular}{c|ccccccccccccccccccccc|cc}
    \hline
        Item& 3VT & BL & 4C & TV & UR & HL & ICO & FL & CTSP & TF & CI & TTV & DI & Ab & LVOT & Kidneys & Eyes & TCV & MFP & LSSP & RVOT & TOP1(\%) & TOP5(\%) \\ \hline
         ResNet18 & 0.96  & 0.88  & 0.77  & 0.61  & 0.68  & 0.86  & 0.94  & 0.73  & 0.96  & 0.73  & \underline{0.85}  & 0.54  & 0.81  & \underline{0.94}  & 0.79  & \underline{0.93}  & 0.95  & \underline{0.98}  & 0.81  & \underline{0.98}  & 0.77 & 82.86 & 99.56 \\ 
         
         ResNet50& 0.93  & 0.85  & 0.78  & 0.66  & 0.60  & \underline{0.88}  & 0.81  & 0.71  & 0.94  & 0.75  & 0.76  & 0.59  & \textbf{1.00}  & 0.83  & 0.79  & \textbf{0.94}  & 0.88  & \textbf{1.00}  & 0.81  & 0.95  & 0.77 & 80.90 & 99.34\\ 
         
         ConvNeXt& \textbf{1.00}  & \underline{0.96}  & 0.94  & 0.62  & 0.66  & 0.68  & 0.93  & 0.77  & 0.87  & 0.72  & 0.75  & 0.60  & 0.88  & 0.86  & 0.70  & 0.84  & 0.93  & 0.96  & 0.89  & \textbf{1.00}  & 0.82 & 82.21 & 99.67\\ 
         
         EfficientNet& 0.93  & 0.87  & 0.63  & \textbf{1.00}  & 0.51  & 0.64  & 0.91  & 0.48  & 0.78  & 0.40  & 0.55  & 0.48  & 0.76  & 0.57  & 0.42  & 0.71  & 0.78  & 0.86  & 0.76  & 0.95   & 0.50 & 65.72 & 97.71\\
         
         MobileNetV2& 0.90  & 0.89  & 0.87  & 0.61  & 0.57  & 0.75  & 0.93  & 0.62  & 0.91  & 0.62  & 0.65  & 0.49  & 0.86  & 0.88  & 0.54  & 0.78  & 0.84  & 0.93  & 0.85  & 0.91  & 0.66 & 75.22 & 99.02\\
         
        VGG16 & \textbf{1.00}  & 0.90  & 0.86  & 0.63  & 0.65  & 0.78  & \underline{0.96}  & 0.81  & \underline{0.98}  & 0.65  & 0.78  & 0.65  & \underline{0.96}  & 0.94  & \textbf{0.81}  & \underline{0.93}  & \textbf{0.98}  & 0.96  & \underline{0.96}  & \underline{0.98}   & 0.86 & 85.48 & \underline{99.78}\\ 
        
         Conformer& 0.96  & 0.91  & 0.83  & 0.76  & \underline{0.78}  & \textbf{0.89}  & \underline{0.96}  & \underline{0.89}  & \underline{0.98}  & \underline{0.76}  & 0.80  & \underline{0.73}  & \textbf{1.00}  & 0.92  & 0.79  & \underline{0.93}  & \underline{0.96}  & \textbf{1.00}  & \underline{0.96}  & \underline{0.98}  & \underline{0.88} & \underline{88.54} & 99.56\\
         
         MvitV2& \textbf{1.00}  & 0.90  & \underline{0.90}  & 0.68  & 0.72  & 0.82  & \textbf{0.98}  & \textbf{0.91}  & \textbf{1.00}  & 0.69  & \underline{0.85}  & 0.63  & \textbf{1.00}  & 0.92  & \underline{0.80}  & 0.91  & \underline{0.96}  & \textbf{1.00}  & \textbf{0.98}  & \underline{0.98}   & \textbf{0.97} & 87.99 & 99.45\\ 
         
         Vit & 0.89  & 0.88  & 0.68  & 0.45  & 0.50  & 0.57  & 0.89  & 0.47  & 0.82  & 0.36  & 0.57  & 0.45  & 0.39  & 0.79  & 0.55  & 0.71  & 0.52  & 0.91  & 0.52  & 0.73  & 0.74 & 62.01 & 95.52\\
         
         SwtV2 & \underline{0.98}  & 0.93  & 0.87  & 0.53  & 0.67  & 0.66  & \underline{0.96}  & 0.70  & \underline{0.98}  & 0.70  & 0.77  & 0.59  & 0.91  & 0.87  & 0.79  & 0.84  & 0.91  & \textbf{1.00}  & 0.95  & \textbf{1.00}  & 0.85 & 82.64 & 99.56 \\
         
         Ours& \textbf{1.00}  & \underline{0.90}  & \textbf{0.95}  & \underline{0.80}  & \textbf{0.81}  & 0.82  & 0.95  & \textbf{0.91}  & \textbf{1.00}  & \textbf{0.79}  & \textbf{0.88}  & \textbf{0.85}  & \textbf{1.00}  & \textbf{0.98}  & \textbf{0.81}  & \underline{0.93}  & \textbf{0.98}  & \textbf{1.00}  & \textbf{0.98}  & \underline{0.98}  & 0.86 & \textbf{91.05} & \textbf{1.00}\\ \hline
    \end{tabular}
    }
    \label{table4}
    \vspace{-5mm}
\end{table*}

\subsection{Model Ablation Experiments}
In order to fully verify that the models in this paper have strong generalization performance, extensive comparative experiments were conducted on mainstream models such as ResNet and VGG and new classification models proposed in 2022 such as ConFormer and Mvitv2, and the results of the experiments are shown in Table~\ref{table4}. These models are the original models without modifications. Our model improves 8.19\% in Top-1 metrics compared to the ResNet-18 baseline model, while the ResNet-50 baseline model lags behind ResNet-18 in Top-1 metrics. This is because ultrasound images are fundamentally different from traditional images, and there is a large amount of mutual interference in the multi-feature classification task of images, and the ResNet is a unidirectional feature transmission. The reason is that ultrasound images are different from traditional images, and there is a large amount of mutual interference in multi-feature classification tasks. Although this paper's model is not only 2.51\% ahead of the next best model in the Top-1 metric, but also the Top-5 metric is 100\% accurately predicted, the Top-5 metric verifies that this paper's model is able to accurately predict the true category of the image within the first 5 prediction results every time it predicts the image.

In addition, this paper further validates the performance of FPDANet by the average False Negative Rate (FNRs). As shown in Fig.~\ref{FNRs}, Fig.~\ref{FNRs} (a) is the histogram of FNRs between this paper and the mainstream classification models, in which the FNRs value of FPDANet is lower than that of all the mainstream classification models, and is 4.95\% lower than that of the suboptimal model, Conformer model. Fig.~\ref{FNRs}(b) shows the classification results of FPDANet vs. ResNet18, where the FNRs values of FPDANet are better than ResNet18 for all classifications.


\section{CONCLUSIONS}
In this paper, an automatic classification model for fetal ultrasound images is designed based on ResNet backbone network, DAN attention mechanism and bilateral multi-scale information fusion network. The backbone network is utilized to extract the underlying texture features, the feature dependency is enhanced by introducing the DAN mechanism, the interference of Identity block short-circuit connection is filtered, and the multi-scale features are fused by the bilateral multi-scale fusion network to make up for the feature singularity of the backbone network and promote the flow fusion. Extensive experiments are conducted in this paper, which show that the proposed FPDANet has good generalization and robustness.

\bibliographystyle{IEEEtran}
\bibliography{article}

\end{document}